\documentclass[12pt]{article}

\usepackage{textcomp}
\usepackage{chetnew}
\usepackage{tikz}

\newcommand{\CW}{\mathcal{W}}
\newcommand{\CB}{\mathcal{B}}

\newcommand{\CA}{\mathcal{A}}
\newcommand{\CQ}{\mathcal{Q}}

\newcommand{\CD}{\mathcal{D}}
\newcommand{\CC}{\mathcal{C}}
\newcommand{\CO}{\mathcal{O}}
\newcommand{\CT}{\mathcal{T}}

\newcommand{\CI}{\mathcal{I}}
\newcommand{\CN}{\mathcal{N}}
\newcommand{\CS}{\mathcal{S}}
\newcommand{\CM}{\mathcal{M}}

\preprint{EFI-16-05, YITP-16-17}

\title{\vspace*{-16mm}Conformal Manifolds in Four Dimensions \\[2mm] and Chiral Algebras}

\author{Matthew Buican$^{\diamondsuit, 1,2}$ and Takahiro Nishinaka$^{\clubsuit, 3}$
}
\affiliation{$^1$CRST and School of Physics and Astronomy \\ Queen Mary University of London, London E1 4NS, UK\\ \smallskip$^2$Enrico Fermi Institute and Department of Physics\\  The University of Chicago, Chicago, IL 60637, USA\\ \smallskip$^3$Yukawa Institute for Theoretical Physics\\ Kyoto University, Kyoto 606-8502, Japan \emails{$^{\diamondsuit}$m.buican@qmul.ac.uk, $^{\clubsuit}$takahiro.nishinaka@yukawa.kyoto-u.ac.jp}}

\abstract{Any $\CN=2$ superconformal field theory (SCFT) in four dimensions has a sector of operators related to a two-dimensional chiral algebra containing a Virasoro sub-algebra. Moreover, there are well-known examples of isolated SCFTs whose chiral algebra is a Virasoro algebra. In this note, we consider the chiral algebras associated with interacting $\CN=2$ SCFTs possessing  an exactly marginal deformation that can be interpreted as a gauge coupling (i.e., at special points on the resulting conformal manifolds, free gauge fields appear that decouple from isolated SCFT building blocks). At any point on these conformal manifolds, we argue that the associated chiral algebras possess at least three generators. In addition, we show that there are examples of SCFTs realizing such a minimal chiral algebra: they are certain points on the conformal manifold obtained by considering the low-energy limit of type IIB string theory on the three complex-dimensional hypersurface singularity $x_1^3+x_2^3+x_3^3+\alpha x_1x_2x_3+w^2=0$. The associated chiral algebra is the $\CA(6)$ theory of Feigin, Feigin, and Tipunin. As byproducts of our work, we argue that {\bf(i)} a collection of isolated theories can be conformally gauged only if there is a SUSY moduli space associated with the corresponding symmetry current moment maps in each  sector, and {\bf(ii)} $\CN=2$ SCFTs with $a\ge c$ have hidden fermionic symmetries (in the sense of fermionic chiral algebra generators).
}

\date{March 2016}

\begin{document}
\setcounter{tocdepth}{2}

\maketitle
\toc

\section{Introduction}
$\CN=2$ SCFTs in four dimensions have at least two natural sub-sectors: the $\CN=2$ chiral ring (i.e., the ring of operators annihilated by all the anti-chiral Poincar\'e supersymmetries) and the sector of Schur operators. The former controls the Coulomb branch moduli space of vacua (when it exists) and was elucidated over twenty years ago in the groundbreaking work of Seiberg and Witten \cite{Seiberg:1994rs,Seiberg:1994aj}. The later captures a more intricate structure that is related to a chiral algebra in two dimensions \cite{Beem:2013sza}. When there is a Higgs branch, the Schur operators describe its properties, but this sector is also much more general.

Surprisingly, recent work on Argyres-Douglas theories \cite{Buican:2015ina, Buican:2015hsa, Buican:2015tda,Cordova:2015nma,Song:2015wta,Cecotti:2015lab} shows that these two sectors are in fact closely related. Indeed, aspects of the Coulomb branch operator spectrum are secretly encoded in the Schur limit of the superconformal index \cite{Buican:2015hsa} (see also the discussion in \cite{Buican:2016hnq} of the deformation described in \cite{Xie:2016hny}). Moreover, BPS calculations on the Coulomb branch allow one to reproduce the Schur index \cite{Cordova:2015nma,Song:2015wta,Cecotti:2015lab}. 

In this work, we will show another way in which the $\CN=2$ chiral ring is tied to the Schur sector. More precisely, we focus on the subset of the $\CN=2$ chiral ring that controls exactly marginal deformations (i.e., the chiral operators of scaling dimension two). We will argue below that if a theory has an exactly marginal deformation, and it can be interpreted as a gauge coupling (i.e., the conformal manifold is built by conformally gauging global symmetries of isolated SCFT sub-sectors), then the chiral algebra, $\chi$, associated with the theory has at least three generators. We will write this statement as
\eqn{
|\chi|\ge3~,
}[bound]
where the norm indicates the number of generators of the chiral algebra (i.e., the number of operators whose normal-ordered products, along with their derivatives, span the chiral algebra). 

Interestingly, it turns out that the bound in \eqref{bound} can be saturated by actual theories. Indeed, we find one such example: the $\CA(6)$ theory of Feigin, Feigin, and Tipunin \cite{Feigin:2007sp,Feigin:2008sg}, and the associated four-dimensional theory is gotten by, for example, studying the low-energy limit of type IIB string theory on a particular three complex-dimensional hypersurface singularity we will mention below. Alternatively, this theory can be engineered by taking three copies of the $(A_1, A_3)$ theory and gauging a diagonal $SU(2)$.

In the context of certain isolated theories, like the $(A_1, A_{2n})$ theories \cite{Cecotti:2010fi}, it is known that the chiral algebra corresponds to a vacuum module of Virasoro \cite{Rastelli,Cordova:2015nma}, so $|\chi|=1$. Our result is a first hint of some new connections between conformal manifolds and chiral algebras. We will briefly return to these points in the conclusions.

We should also note that our argument in favor of \eqref{bound} has at least two interesting spinoffs:
\begin{itemize}
\item[${\bf(i)}$] We show that we can conformally gauge some diagonal global symmetry, $H$, of a set of isolated theories, $\CT_i$, only if the corresponding holomorphic moment maps, $\mu_i^I$ (with $I=1,\cdots,|H|$ an adjoint index), satisfy $\left(\mu_i^I\right)^n\ne0$ in the chiral ring of the $\CT_i$ for all $n>0$. A standard folk theorem (which we do not prove) then implies that there is an associated SUSY moduli space in each $\CT_i$.
\item[${\bf(ii)}$] Assuming all $\CN=2$ SCFTs (or at least the ones we study) have the asymptotic Cardy-like behavior governed by $a-c$ derived in \cite{DiPietro:2014bca}, then it follows that for any $\CN=2$ theory with $a\ge c$, the associated chiral algebra necessarily has fermionic generators.\foot{In \cite{Buican:2015ina}, we saw that $a-c$ does indeed control the $q\to1$ asymptotics of the Schur index for collections of free hypermultiplets and vector multiplets as well as for several infinite classes of non-Lagrangian theories (see also the recent discussion in \cite{Ardehali:2015bla}) via
\eqn{
\lim_{\beta\to0}\log\CI_{S}(q)=-{8\pi^2\over\beta}(a-c)+\cdots~.
}[Schurcardy]
In this formula, $\CI_S(q)$ is the Schur index, and $q=e^{-\beta}$ is the corresponding superconformal fugacity (see the discussion in section \ref{basics} for further details). On the other hand, for certain $\CN=1$ theories, \cite{Ardehali:2015bla} argues that the Cardy-like behavior of \cite{DiPietro:2014bca} is modified. We assume that the theories we study have the standard Cardy-like behavior controlled by $a-c$ in \eqref{Schurcardy} (and we are not aware of any $\CN=2$ SCFT counterexamples to this behavior).} This result implies that $|\chi|\ge3$ for interacting theories with $a\ge c$.
\end{itemize}

The first point indicates a surprising connection between two different spaces: the conformal manifold on the one hand and the moduli space of vacua on the other (in all known examples, there is a more obvious connection between the conformal manifold and the Coulomb branch, because the exactly marginal deformations sit in multiplets whose primaries parameterize part of the Coulomb branch).\foot{Connections between conformal manifolds and moduli spaces of vacua arise often in the context of the AdS/CFT correspondence. However, the relations we find are purely in the realm of field theory.} The second point is interesting as well since it implies that $\CN=2$ theories with $a\ge c$ have, in some sense, additional fermionic symmetries. In the case of theories with $\CN>2$ (which necessarily have $a=c$), these additional symmetries include additional supersymmetries in four space-time dimensions (and therefore additional supersymmetries in two-dimensions---see \cite{Nishinaka:2016hbw} for a recent discussion of the case with 4D $\CN=3$ and \cite{Beem:2013sza} for the case with 4D $\CN=4$). On the other hand, in the case of theories with only $\CN=2$ SUSY, these additional fermionic symmetries only exist at the level of the corresponding chiral algebras.

The plan of this paper is as follows. In the next section, we review some chiral algebra and Schur sector basics. We then describe conformal gauging and what it means for the Schur sector. Afterwards, we move on to describe some empirical evidence in favor of \eqref{bound}. Section 4 is the meat of the paper and contains our physics proof of \eqref{bound}. In section 5, we then describe our realization of a theory that saturates \eqref{bound}. Finally, we end with conclusions and a discussion of open problems.

\section{Chiral algebra / Schur sector basics}
\label{basics}
In this section we will give a lightning review of the pertinent aspects of Schur operators \cite{Gadde:2011uv} and their associated chiral algebras \cite{Beem:2013sza} that are useful below. For further details, the reader is encouraged to consult the original references.

The Schur sector consists of the operators that are annihilated by the two supercharges $\tilde\CQ_{2\dot-}$ and $\CQ^1_-$ (along with their conjugates). These degrees of freedom are the operators that contribute to the Schur limit of the superconformal index
\begin{align}
\mathcal{I}(q;\vec{x}) = \text{Tr}_{\mathcal{H}}(-1)^F e^{-\beta\Delta}q^{E-R}\prod_{i}(x_i)^{f_i}~.
\label{Schurlim}
\end{align}
Here $|q|<1$, the $|x_i|=1$ are flavor symmetry fugacities, $E$ is the scaling dimension, $R$ is the $SU(2)_R$ Cartan, and $\Delta=\left\{\CQ_{2\dot-},(\CQ_{2\dot-})^{\dagger}\right\}$.
In the classification of \cite{Dolan:2002zh}, the multiplets that contain Schur operators are of type (see also the original discussion in \cite{Dobrev:1985qv})
\eqn{
\hat\CB_R~,  \ \ \ \CD_{R(0,j_2)}\oplus\bar\CD_{R(j_1,0)}~, \ \ \ \hat\CC_{R(j_1,j_2)}~,
}[Schurops]
where the $j_i$ are the Lorentz spins.

It turns out that the $\hat\CB_R$ and $\bar\CD_{R(j_1,0)}$ form a subring called the Hall-Littlewood (HL) chiral ring, and it will be important for us below. As a point of reference, the $\hat\CB_R$ multiplets contain Schur operators that are the highest-weight primaries and parameterize the Higgs branch (when it exists). In particular, $\hat\CB_1$ is the multiplet corresponding to flavor symmetry currents, and the Schur operator is the holomorphic moment map, $\mu$. On the other hand, the $\CD_{R(0,j_2)}\oplus\bar\CD_{R(j_1,0)}$ multiplets contain, respectively, first level $\tilde \CQ_{2\dot+}$ and $\CQ^1_+$ highest-weight descendants as Schur operators. When $R=j_1=j_2=0$, these multiplets are vector multiplets, and the corresponding Schur operators are gauginos, $\lambda^1_+, \tilde\lambda_{2\dot+}$. The $\hat\CC_{R(j_1,j_2)}$ are semi-short multiplets and contain Schur operators as $\tilde \CQ_{2\dot+}\CQ^1_+$ highest-weight descendants. The $\hat\CC_{0(0,0)}$ multiplet is the stress tensor multiplet, and the corresponding Schur operator is the highest weight state of the $SU(2)_R$ current, $J^{11}_{+\dot+}$ (where the superscripts are $SU(2)_R$ spinor indices).

The authors of a beautiful recent paper \cite{Beem:2013sza} found a deep connection between the above Schur operators and two-dimensional chiral algebras. Roughly speaking, one fixes the Schur operators to lie in a plane and works with respect to a cohomology defined by a particular nilpotent supercharge. The Schur operators form non-trivial representatives of this cohomology. Moreover, the Schur operators behave meromorphically (i.e., their correlation functions in the plane are meromorphic) as long as we twist the anti-holomorphic translations in the plane with the $SU(2)_R$ symmetry. Therefore, twisted-translated cohomology classes corresponding to Schur operators map to chiral algebra operators.

As an illustration of the above discussion, we have the following natural 4d/2d maps \rcite{Beem:2013sza}
\eqn{
\chi\left[J_{+\dot+}^{11}\right]=-{1\over2\pi^2}T~, \ \ \ \chi\left[\mu^{I}\right]={1\over2\sqrt{2}\pi^2}J^I ~, \ \ \chi\left[\partial_{+\dot+}\right]=\partial_z\equiv\partial~,
}[Univdictionary]
where $\chi[\cdots]$ denotes the chiral algebra image of the argument, $T$ is the holomorphic stress tensor, $J^I$ is an affine current, and $\partial$ is a holomorphic derivative. Moreover, the central charges in four dimensions map simply to corresponding charges in two dimensions: $c_{2d}=-12c_{4d}$ and $k_{2d}=-{1\over2}k_{4d}$. In addition, the holomorphic level, $h$, satisfies
\eqn{
h=E-R~.
}[hvsER]
We can write the torus partition function of the chiral algebra as follows
\eqn{
Z(x,q)={\rm Tr}\ x^{M^{\perp}}q^{L_0}~,
}[torus]
where $h$ is the eigenvalue of $L_0$, and the trace is taken over the chiral algebra states. In \eqref{torus}, $x$ is a fugacity for the rotations in the plane transverse to the chiral algebra plane. From the perspective of the chiral algebra, this is an additional symmetry, and it will play an important role in our example below. Setting $x=-1$ and using the fact that spin-statistics implies that $(-1)^{M^{\perp}}=(-1)^F$, where $F$ is the fermion number, it is straightforward to argue that the torus partition function should match the Schur index in \eqref{Schurlim}
\eqn{
Z(-1,q)=\CI(q)~.
}[torusI]
We will use this correspondence below to identify the $\CA(6)$ theory with the chiral algebra of our diagonally gauged AD theory.

The important points of the above chiral algebra story for us are the following
\begin{itemize}
\item[$\bullet$] From \eqref{Univdictionary}, we see that the two-dimensional chiral algebra {\it always} has a Virasoro sub algebra since there is always an $SU(2)_R$ current in four dimensions. In the next few sections, we will see what additional structure the conformal manifold forces on this algebra.
\item[$\bullet$] The generators of the HL ring are automatically chiral algebra generators \cite{Beem:2013sza}.
\item[$\bullet$] If the four-dimensional theory has flavor symmetry moment maps that vanish (in the chiral ring) at second order in the singlet channel, then $T$ in \eqref{Univdictionary} is not an independent generator. Instead, it is the Sugawara stress tensor.
\item[$\bullet$] Higgs branch generators contribute at least two chiral algebra generators since the Higgs branch is quaternionic. Moreover, interacting theories with Higgs branches necessarily have $|\chi|\ge3$ since the stress tensor typically contributes an additional generator. In order for the stress tensor to be a Sugawara composite in two dimensions, there must be at least three $\hat\CB_R$ generators (and a corresponding number of generators in two dimensions) \cite{Beem:2013sza}.
\item[$\bullet$] The previous point also applies to $\hat\CB_R$ generators whose vevs parameterize more general moduli spaces where $SU(2)_R$ is broken but $U(1)_R$ is preserved. For example, these moduli spaces can contain some number of free vector multiplets in addition to free hypermultiplets at generic points.
\end{itemize}

\subsection{Gauging}
\label{Gauging}

In this section we suppose that we start with a collection of $N$ isolated SCFTs and gauge a diagonal $H$ symmetry with coupling $g$
\eqn{
\delta W=g\cdot\Phi\sum_{i=1}^N\mu_i~,
}[WdefSU(2)diagonal]
where $\Phi$ is the adjoint field, and the $\mu_i$ are the $H$ moment maps of the different isolated sectors, $\CT_i$. We then want to know what happens to the chiral algebra.

In very general terms, when we turn on a non-zero gauge coupling, many of the multiplets that were short at $g=0$ pair up and become long multiplets (according to the rules of \cite{Dolan:2002zh}). As a result, the chiral algebra that appears at infinite Zamolodchikov distance when $g=0$ typically becomes much smaller when we are at finite Zamolodchikov distance and non-zero gauge coupling. For operators built out of the Schur gauginos and moment maps, this pairing up can be explicitly worked out at small gauge coupling\foot{Note that this regime does not require the isolated SCFTs to be weakly coupled. Indeed, the most interesting examples to study (including the one we will discuss in more detail below) involve strongly coupled isolated sectors.} via the following (anti)commutators
\begin{eqnarray}\label{lambdadef}
\left\{\tilde Q_{2\dot-}, \tilde\lambda_{2\dot+}\right\}&=&\left\{Q^1_-, \lambda_+^1\right\}=F=g\sum_{i=1}^N\mu_i~,\ \ \ \left\{\tilde Q_{2\dot-},\lambda^1_+\right\}=\left\{Q^1_-,\tilde\lambda_{2\dot+}\right\}=0~,\nonumber\\ \left[Q^1_-,D_{+\dot+}\right]&=&g\tilde\lambda^1_{\dot+}=g\tilde\lambda_{2\dot+}~, \ \ \ \left[\tilde Q_{2\dot-},D_{+\dot+}\right]=g\lambda^1_{+}~, \ \ \ \left[Q^1_-,D_{+\dot+}\right]=g\tilde\lambda^1_{\dot+}=g\tilde\lambda_{2\dot+}~,\nonumber\\  \left[\tilde Q_{2\dot-},D_{+\dot+}\right]&=&g\lambda^1_{+}~,\ \ \ \left[Q^1_-,\mu_I\right]=\left[\tilde Q_{2\dot-},\mu_I\right]=0~.
\end{eqnarray}
The last two commutators amount to imposing current conservation. Most of the operators built out of these degrees of freedom pair up at non-zero gauge coupling.

The important points for us below are that 
\begin{itemize}
\item[$\bullet$] Interacting theories with fermionic chiral algebra generators must have $|\chi|\ge3$ since these operators come in pairs and cannot form a stress tensor composite.
\item[$\bullet$] We do not require that the chiral algebra is invariant as we explore finite diameter subregions of the conformal manifold (although there is considerable evidence that in many theories this invariance holds).
\end{itemize}

\section{Phenomenology}
Before we get to our argument in favor of $|\chi|\ge3$ in the case of theories with a marginal gauge coupling, let us review some of the strong phenomenological and empirical evidence in favor of this proposition.

For example, $SU(N_c)$ SQCD with $N_f=2N_c\ge4$ has at least $U(2N_c)$ flavor symmetry. Therefore, the associated chiral algebra at any point in the coupling space must have at least $4N_c^2\ge16$ generators.\foot{In the simplest case of $SU(2)$ with $N_f=4$, the global symmetry group is enhanced to $SO(8)$, and the authors of \cite{Beem:2013sza} have conjectured that the chiral algebra away from decoupling points is the $so(8)$ Affine Kac-Moody algebra at level $k_{2d}=-2$. For $N_c>2$, the authors conjecture that the chiral algebra is generated by affine $u(2N_c)$ currents at level $k_{2d}=-N_c$ along with baryonic generators $b$ and $\tilde b$ obeying a particular OPE.} Similar statements hold in the case of superconformal SQCD with $SO$ and $Sp$ gauge groups.

Moving to the more general class $\CS$ context with regular punctures, we can typically reduce the flavor symmetry,\foot{Since flavor symmetries must contribute generators to the chiral algebra, reducing the flavor symmetry can lead to smaller $|\chi|$.} and, in the process, induce exactly marginal deformations by going to higher genus Riemann surfaces. On the other hand, quite generally, higher genus surfaces lead to theories with $\CD\oplus\bar\CD$ representations \cite{Gadde:2011uv}. As discussed above, this situation leads to theories with $|\chi|\ge3$. For example, the authors of \cite{Beem:2013sza} studied the chiral algebra associated with a genus two $SU(2)^3$ gauge theory. The flavor symmetry is reduced to $U(1)$, but there are many $\CD\oplus\bar\CD$ representations, and the chiral algebra has $|\chi|>3$. 

Finally, let us consider the class of (generalized) Argyres-Douglas theories. Although this is an enormous zoo, we can say some things about the associated chiral algebras. For simplicity, let us stick to the set of $(A_{K-1}, A_{N-1})$ theories \cite{Cecotti:2010fi}. Such theories have marginal couplings only if $K$ and $N$ are not relatively prime.\foot{Note that this is not a biconditional statement. For example, the $(A_1, A_{2n-3})$ theories discussed in \cite{Buican:2015ina, Buican:2015hsa, Buican:2015tda} are isolated.} These theories have (often large) Higgs branches. Indeed, the three-dimensional mirror analysis of \cite{Xie:2012hs,Xie:2013jc} shows that such theories have a $2(K-1)\ge2$ complex dimensional Higgs branch. It follows from one of our bullet points in the first part of section \ref{basics} that $|\chi|\ge3$. As a particular set of examples of such theories with a marginal coupling, we can consider the $(A_N, A_N)$ theories for $N\ge3$ (the conformal manifold of the $N=3$ case was explored in \cite{Buican:2014hfa, DelZotto:2015rca}). By our above discussion, $|\chi|> 2N\ge6$.

\section{An argument in favor of $|\chi|\ge3$}
Below we give a detailed argument explaining why the inequality in \eqref{bound} should be satisfied in theories with an exactly marginal gauge coupling.

\subsec{The case of $N\ge4$}
When the conformal manifold has regions in which $N\ge4$ isolated SCFTs, $\CT_i$ ($i=1,\cdots, N$), weakly coupled to a gauge field appear (we denote $g=0$ as a strict decoupling point), it is straightforward to see that  $|\chi|\ge3$ by a direct argument in superconformal representation theory.\foot{In this sub-section, we will also assume that each sector has a unique flavor-neutral $\hat\CC_{0(0,0)}$ multiplet. In the more general argument of the next subsection, we will abandon this assumption.} Indeed, consider the Schur conformal primary operators at $\CO(q^2)$. On general grounds, these operators must be in multiplets of type $\hat\CC_{0(0,0)}$, $\hat\CB_2$, $\CD_{0(0,1)}\oplus\bar\CD_{0(1,0)}$, $\CD_{1(0,0)}\oplus\bar\CD_{1(0,0)}$, or $\CD_{{1\over2}(0,{1\over2})}\oplus\bar\CD_{{1\over2}({1\over2},0)}$ \cite{Gadde:2011uv}. 

For the purposes of our argument, it suffices to focus on the  $\hat\CC_{0(0,0)}$, $\hat\CB_2$, and $\CD_{1(0,0)}\oplus\bar\CD_{1(0,0)}$ multiplets. These degrees of freedom can only recombine into long multiplets via \cite{Dolan:2002zh}
\eqn{
\hat\CC_{0(0,0)}\oplus\CD_{1(0,0)}\oplus\bar\CD_{1(0,0)}\oplus\hat\CB_2~.
}[multrecomnew]
Clearly, the $N(N-1)/2$ flavor-singlet operators $\CO_{ab}=\mu_a^I\mu_{bI}\in\hat\CB_2$ with $a\ne b$ are in short multiplets when $g=0$ (here $I=1,\cdots,|H|$ is an adjoint index). By \eqref{multrecomnew}, these multiplets can only recombine if there are an equal number of (flavor-singlet) $\hat\CC_{0(0,0)}$ multiplets. These latter multiplets contain a stress tensor, and there are $N$ such multiplets available to pair up with the $\CO_{ab}$ multiplets when we turn the gauge coupling on. As a result, there are $N(N-3)/2\ge 2$ unpaired $\CO_{ab}$ multiplets. In particular, it follows that $|\chi|\ge3$ as promised. Note that this is a strict lower bound at any point on the conformal manifold since these multiplets can disappear from the short sector only by pairing up with new conserved stress tensor multiplets (which come with an equal number of short $\hat\CB_2$ and $\CD_{1(0,0)}\oplus\bar\CD_{1(0,0)}$ multiplets).

Before we conclude this sub-section, let us discuss some subtleties related to free fields. To that end, we suppose, as part of our collection of isolated theories at $g=0$, we have free hypermultiplets transforming in $n_1\ge1$ copies of a representation, $R_1$, of $H$. In this case, we define the collection of $n_1$ copies of $R_1$ to form a single isolated sector, $\CT_1$. The reason we do not count each of the $n_1$ copies of $R_1$ as forming different sectors is that the corresponding flavor singlet $\hat\CC_{0(0,0)}$ multiplet used in our argument above involves multiplets in all $n_1$ copies of $R_1$.

To understand this statement, consider $H=SU(2)$ SCQD with $N_f=4$. We have eight fundamental $SU(2)$ doublets, $Q_{a}^{Ai}$, and an $SO(8)$ flavor symmetry (here $A=1,2$ is an $SU(2)_R$ index, $a=1,2$ is a fundamental $H$ gauge index, and $i=1,\cdots,8$ is a fundamental $SO(8)$ index). This theory has thirty-seven short $\hat\CC_{0(0,0)}$ multiplets at $g=0$
\eqn{
J_{\lambda}\sim\Phi^I\Phi^{\dagger}_I~, \ \ \ J^{ij}\sim\epsilon_{AB}\epsilon^{ab}Q^{Ai}_{a}Q^{Bj}_{b}~,
}[su2short]
where $\Phi^I$ is the chiral primary of the $SU(2)$ vector multiplet. The only flavor singlet $\hat\CC_{0(0,0)}$ operator arising from the hypermultiplet sector is $J_m=\sum_{i=1}^8J^{ii}$, and so it is sensible to define the collection of $n_1$ copies of $R_1$ as a single sector.

Note that when we turn on $g\ne0$, the only remaining short operator in this collection is
\eqn{
J=J_{\lambda}-J_m~.
}[su2shortg]
The other thirty-six short $\hat\CC_{0(0,0)}$ multiplets pair up with an equal number of $\hat\CB_2$ multiplets. In particular, consider the flavor singlet operator
\eqn{
J_{\perp}=\kappa_{\lambda}J_{\lambda}+J_m~,
}[perpcurr]
where $\kappa_{\lambda}\ne0$ is a constant chosen so that $\langle J_{\perp}(x)J(0)\rangle=0$ at $g=0$. This operator pairs up with the flavor singlet $\mu^{ij}\mu_{ij}$ operator to form a long multiplet, where
\eqn{
\mu^{ij}=\epsilon^{ab}Q^{1i}_{a}Q^{1j}_{b}~.
}[defnmu]
When we turn on small $g\ne0$, this statement implies
\eqn{
\mu^{ij}\mu_{ij}=0~,
}[chirringrel]
at the level of the chiral ring, while, at the level of the $\mu^{ij}(x)\mu_{ij}(0)$ OPE, we have
\eqn{
\mu^{ij}(x)\mu_{ij}(0)=C\cdot x^{\Delta-4}\cdot(Q^1)^2(\tilde Q_2)^2J_{\perp}+\cdots~,
}[mumuOPE]
where $\Delta>4$. In particular, the term appearing on the RHS of \eqref{mumuOPE} arises because $\mu^{ij}\mu_{ij}$ pairs up with $J_{\perp}$.

\subsec{The case of $N=1$}
The above argument does not work when the $g=0$ point consists of a decoupled gauge multiplet and a single isolated SCFT sector, $\CT$. Instead, we will take a more indirect route in this case (note that, as we will see in the next sub-section, the argument in this sub-section generalizes to arbitrary numbers of sectors; moreover, we do not assume that each isolated SCFT has a single flavor-singlet $\hat\CC_{0(0,0)}$ multiplet; instead we allow for one $\hat\CC_{0(0,0)}$ multiplet that contains the genuine stress tensor of the sector---i.e., the operator that gives rise to the generator of translations in the sector---and an arbitrary number of other $\hat\CC_{0(0,0)}$ multiplets). To that end, we first show that $\CT$ has a moduli (sub)space parameterized by vevs of the holomorphic moment maps. From this fact, we will be able to show that if $|\chi|<3$, then $a_{4d}\ge c_{4d}$ (note that we add the subscript \lq\lq4d" to $a$ solely for unity of notation)  along the conformal manifold of the gauged theory. Next, assuming the behavior described in \cite{DiPietro:2014bca} and \eqref{Schurcardy} (see the introduction for a discussion of this point), we see that the partition function of the resulting chiral algebra lacks an essential singularity in the $q\to1$ limit. Finally, we argue that the absence of this singularity implies the existence of fermionic chiral algebra generators and hence, by one of the bullet points in section \ref{Gauging}, $|\chi|\ge3$.

We begin by proving that in $\CT$, the $\CO^n=(\mu^I\mu_I)^n\in\hat\CB_{2n}$ operators transform in short multiplets for all $n>0$. In other words, we prove that $\CO^n\neq 0$ in the chiral ring of $\CT$. For $n=1$, this follows from the unitarity bound of \cite{Beem:2013sza}. Indeed, if it were not the case, then (4.16) of \cite{Beem:2013sza} implies
\eqn{
{{\rm dim} H\over c_{4d}}={24 h^{\lor}\over k_{4d}}-12=-6~,
}[kcbd]
and so $c_{4d}<0$, which contradicts four-dimensional unitarity. More generally, suppose that there is some maximal $\hat n\ge1$ such that $\CO^{\hat n-1}$ is in a short multiplet of $\CT$. As a result, $\CO^{\hat n}$ is in a long multiplet of type\foot{We will see below that it is also inconsistent for $\CO^{\hat n}$ to identically vanish.}
\eqn{
\hat\CC_{2\hat n-2(0,0)}\oplus\CD_{2\hat n-1(0,0)}\oplus\bar\CD_{2\hat n-1(0,0)}\oplus\hat\CB_{2\hat n}~.
}[hatnlong]
Let us denote the (would be) Schur operators in the first three sub-multiplets of \eqref{hatnlong} as $\CO'_{+\dot+}$, $\CO''_{\dot+}$, and $\CO''_{+}$ respectively.

Now, consider working at arbitrarily small $g\ne0$. The theory still possesses a short stress tensor multiplet, and this multiplet has an associated Schur operator of the form
\eqn{
J^{11}_{+\dot+}=(\lambda^1_+)^I(\tilde\lambda_{2\dot+})_I-J^{11}_{1+\dot+}~,
}[shortT]
where, at $g=0$, $J^{11}_{1+\dot+}$ is the Schur operator in the stress tensor multiplet of $\CT$. Given this operator, let us define
\eqn{
\tilde\CO_{+\dot+}=\tilde\kappa_{\lambda}(\lambda^1_+)^I(\tilde\lambda_{2\dot+})_I+J^{11}_{1+\dot+}~,
}[orthlong]
where $\tilde\kappa_{\lambda}\ne0,-1$ is a constant chosen so that $\langle J_{+\dot+}^{11}(x)\tilde\CO_{+\dot+}(0)\rangle=0$ (at $g=0$). Clearly, $\tilde\CO_{+\dot+}$ is in a long multiplet for small $g\ne0$ since it pairs up as follows
\eqn{
\tilde\CO_{+\dot+}\oplus\tilde\CO_+\oplus\tilde\CO_{\dot+}\oplus(1+\kappa_{\lambda})\CO~,
}[longiso]
where $\CO_{+}=(1+\kappa_{\lambda})\mu^I(\lambda^1_{+})_I$, $\CO_{\dot+}=(1+\kappa_{\lambda})\mu^I(\tilde\lambda_{2\dot+})_I$, and $\CO=\mu^I\mu_I$. Moreover, we have that
\eqn{
\CO^{\hat n-1}\tilde\CO_{+\dot+}\oplus\CO^{\hat n-1}\tilde\CO_+\oplus\CO^{\hat n-1}\tilde\CO_{\dot+}\oplus(1+\kappa_{\lambda})\CO^{\hat n}~.
}[shortlongcont]
It then follows that $\CO'_{+\dot+}-(1+\kappa_{\lambda})^{-1}\CO^{\hat n-1}\tilde\CO_{+\dot+}$ is in a short multiplet for small $g\ne0$. However, this statement leads to a contradiction, because, by dialing $g$ arbitrarily small but non-zero, we find that the anomalous dimension of $\CO^{\hat n-1}\tilde\CO_{+\dot+}$ is parametrically smaller than the anomalous dimension of $\CO'_{+\dot+}$.\foot{This argument can be suitably generalized if the operators in question do not have definite scaling dimension. Indeed, since the $\CO^{\hat n-1}\tilde\CO_{+\dot+}$ operator must become part of a short multiplet as the gauge coupling is taken to zero, $\CO'_{+\dot+}-\CO^{\hat n-1}\tilde\CO_{+\dot+}$ cannot be part of a short multiplet for sufficiently small $g$.} As a result, we see that, as claimed,
\eqn{
\CO^n\ne0\ \ \ \forall n>0~,
}[nonnilpotent]
in the chiral ring of the isolated theory, $\CT$.\foot{This result immediately implies (after invoking flavor covariance) that
\eqn{
\left(\mu^I\right)^n\ne0\ \ \ \forall n>0,I~,
}[nonnilpotentii]
in the chiral ring of $\CT$.}

As a final aside, note that, a priori, we might have imagined that $\CO^{\hat n}$ could be {\it identically zero}, i.e., it might not even be possible to define this operator to be in a long multiplet as in \eqref{hatnlong}. However, such a statement would be in contradiction with \eqref{shortlongcont}, since $\tilde\CO_{+\dot+}$ must recombine to become part of a long multiplet for $g\ne0$, and therefore $\CO^{\hat n-1}\tilde\CO_{+\dot+}$ should also be part of a long multiplet at small (but non-zero) coupling (in particular, this statement must hold when we take the leading corrections in $g$ into account).

Now we must use some physical intuition. Indeed, since $\CO$ is an $SU(2)_R$-charged but $U(1)_R$-neutral chiral operator that is not nilpotent, it is reasonable to associate a SUSY moduli space, $\hat{\CM}_{\CO}\subset\CM_{SU(2)_R}$, with its vevs (here $\CM_{SU(2)_R}$ contains any normal directions to generic points on $\hat\CM_{\CO}$ that also break $SU(2)_R$ but preserve $U(1)_R$). Given this picture, we expect that for generic $\langle\CO\rangle\ne0$, the IR theory is just a collection of free hypermultiplets. In some theories, these free hypermultiplets might necessarily be accompanied by free vector multiplets as well (see, e.g., \cite{Hanany:2010qu}). This insight allows us to compute a bound on the value of $a_{4d}-c_{4d}$ for the isolated theory
\eqn{
a_{4d, \CT}-c_{4d, \CT}\ge-{1\over24}{\rm dim}\CM_{SU(2)_R}~.
}[amcisoT]
Note that \eqref{amcisoT} follows from the definition of our moduli space: it is a space of vacua in which {\bf(i)} $SU(2)_R$ is spontaneously broken, but $U(1)_R$ is not, and {\bf(ii)} at generic points, the theory consists of a set of free hypermultiplets possibly with additional vector multiplets. As a result, \eqref{amcisoT} follows from linear $U(1)_R$ 't Hooft anomaly matching. In particular, the above inequality is saturated if we have a genuine Higgs branch (i.e., no vector multiplets at generic points).

We can now show that, in the gauged theory, either $a_{4d}\ge c_{4d}$ along the conformal manifold, or $|\chi|\ge3$. Indeed, suppose $a_{4d}<c_{4d}$. In this case, the gauged theory would still possess a non-trivial moduli space of the type described above with $\hat\CB_R$ generators parameterizing it. As discussed in the bullet points of the first part of section \ref{basics}, this conclusion means we would have $|\chi|\ge3$. In particular, if we would like a theory with $|\chi|<3$, we must have that
\eqn{
a_{4d}\ge c_{4d}~,
}[accond]
 along the conformal manifold.

 In the rest of this sub-section, we show that $|\chi| \ge 3$ even in the case of $a_{4d}\ge c_{4d}$.
Note first that, assuming the partition function has the behavior described in \cite{DiPietro:2014bca} and \eqref{Schurcardy}, it is straightforward to show that if $a_{4d}\geq c_{4d}$ then the Schur index lacks an essential singularity as $q\to1$. Let us see if we can find a chiral algebra with $|\chi|=1$ that is compatible with this asymptotic behavior. Since the stress tensor always exists, such an algebra must be a Virasoro algebra. If there are no null vectors (besides the trivial one involving the derivative of the identity) in the vacuum Virasoro module, then the partition function is
\eqn{
\CI=\chi_{(1,1)}={\rm P.E.}\left({q^2\over1-q}\right)~,
}[norelat]
where the plethystic exponential of a function of variables $x_1, \cdots, x_i$ is defined as
\eqn{
P.E.(f(x_1, \cdots, x_i))\equiv {\rm exp}\left(\sum_{n=1}^{\infty}n^{-1}f(x_1^n,\cdots,x_i^n)\right)~.
}[PEdefn]
Therefore, \eqref{norelat} clearly has an essential singularity as $q\to1$. This result implies that the corresponding Schur index also has an essentail singularity and therefore $a_{4d}<c_{4d}$. 

Let us then consider cases with non-trivial null vectors.
From the general formula for the Kac determinant in (7.28) of \cite{DiFrancesco:1997nk} with $h=0$ (since we are considering the Virasoro vacuum representation) as well as (7.31) of that same reference, we find that a non-trivial null vector exists only if
\eqn{
c_{2d}=13-6(t+t^{-1})~, \ \ \ t={1\pm s\over 1\pm r}={p\over p'}>{3\over2} \ {\rm or}\ t<{2\over3}~, 
}[difrancconseq]
for integers $r$, $s$, $p$, and $p'$. Without loss of generality, we can take $p,p'>1$ relatively prime. Note that if $p$ and $p'$ have different sign, then $c_{2d}>0$ and $c_{4d}<0$, which violates four-dimensional unitarity. Also, if $p=1$ or $p'=1$, then we have a partition function as in \eqref{norelat}. Finally, for $t={3\over2}$ or $t={2\over3}$, we have that $c_{2d}=0$ and so by our discussion below \eqref{Univdictionary}, $c_{4d}=0$ as well. However, this statement violates four-dimensional unitarity, so we need not consider this case.

Now, according to (8.17) of \cite{DiFrancesco:1997nk} (we set the overall normalization of the character so that the vacuum contributes unity) we have
\eqn{
\CI=\chi_{(1,1)}={q^{-(p-p')^2/4pp'}\over(q;q)}\sum_{n\in{\bf Z}}\left(q^{(2pp'n+p-p')^2/4pp'}-q^{(2pp'n+p+p')^2/4pp'}\right)~,
}[vacchar]
From the modular transformation properties of the vacuum character, we see that the Schur index \eqref{vacchar} behaves in the limit $q \equiv e^{-\beta} \to 1$ as
\eqn{
\CI\sim e^{-\frac{4\pi^2}{\beta} (h_{\text{min}}-\frac{c_{2d}}{24}) + \mathcal{O}(1)} 
=e^{{\pi^2\over6\beta}\left(1-{6\over pp'}\right)+\CO(1)}~,
}[esssing]
where $h_{\text{min}} = \frac{1-(p-p')^2}{4pp'}$ is the smallest holomorphic dimension of a primary state in the $(p,p')$ minimal model (see \cite{Cecotti:2015lab} for a recent discussion of consequences of this fact in the context of 4D/2D relations).
Since $p$ and $p'$ are relatively prime and $t$ is in the range described in \eqref{difrancconseq}, we see that $pp'\ge10$ ($pp'=10$ in the case of the Yang-Lee model that is related to the $(A_1, A_2)$ theory). Therefore, we see that there is always an essential singularity in \eqref{esssing} and so $a_{4d}<c_{4d}$. This inequality means that the four-dimensional theory cannot have $a_{4d}\geq c_{4d}$ if the corresponding chiral algebra has only one generator.\footnote{See \cite{Cordova:2015nma, Rastelli} for a discussion of four-dimensional $\CN=2$ SCFTs whose chiral algebras are vacuum modules of Virasoro minimal models. All of these theories have $a_{4d}<c_{4d}$.}

We therefore require at least one more generator to cancel the essential singularity in the Schur index and realize $a_{4d}\geq c_{4d}$. However, if there are only two generators, one of them is the stress tensor and the other must be a bosonic operator. Since adding a bosonic generator does not cancel the above essential singularity in the vacuum Virasoro character, we conclude that $a_{4d}<c_{4d}$ even in the case of $|\chi|=2$. Therefore, we need three or more generators of the chiral algebra (including fermions) for the four-dimensional theory to have $a_{4d}\geq c_{4d}$.

\subsec{The case of $N\ge2$}
Let us now study the case of $N\ge2$ isolated SCFT sectors, $\CT_{i}$ ($i=1,\cdots, N$), at $g=0$ with decoupled gauge fields for a diagonal $H$ global symmetry of the $\CT_i$. We will generalize the discussion of the previous subsection for $N=1$. To that end, consider the $N$ operators $\CO_i=\mu^I_i\left(\sum_{a=1}^N\mu_{aI}\right)$. We claim that, in the chiral ring of the collection of the isolated $\CT_i$
\eqn{
\CO_i^n\ne0~,\ \ \ \forall n>0, i=1,\cdots, N~.
}[nonnilp]
From this result, it follows (as in the $N=1$ case discussed previously) that each of the isolated $\CT_i$ SCFTs have moduli spaces, $\CM_{SU(2)_R}$, where $SU(2)_R$ is spontaneously broken but $U(1)_R$ is not.\foot{Moreover, we have
\eqn{
\left(\mu^I_i\right)^n\ne0, \ \ \ \forall n>0, i=1,\cdots, N, I~,
}[flavalli]
in the appropriate chiral rings of the isolated $\CT_i$.} The argument we used in the single sector case starting with \eqref{amcisoT} (but now for all $N$ isolated sectors) applies directly to the case of $N$ sectors since we have a non-trivial $\CM_{SU(2)_R}$. We again conclude that the chiral algebra must have at least three generators. 

To justify \eqref{nonnilp}, we first note that when we turn on small $g\ne0$, only one stress tensor multiplet remains short. This multiplet has the following Schur operator
\eqn{
J^{11}_{+\dot+}=(\lambda^1_+)^I(\tilde\lambda_{2\dot+})_I-\sum_{i=1}^NJ^{11}_{i+\dot+}~,
}[short2]
where (at $g=0$) the $J^{11}_{i+\dot+}$ are the Schur operators in the stress tensor multiplets of the isolated $\CT_i$ theories. We can form $N$ additional linearly independent combinations of the operators appearing on the RHS of \eqref{short2}
\eqn{
\tilde\CO_{a+\dot+}=\kappa_{a,\lambda}(\lambda^1_+)^I(\tilde\lambda_{2\dot+})_I+\sum_{i=1}^N\kappa_{a,i}J^{11}_{i+\dot+}~,
}[ortholincomb]
where $a=1,\cdots,N$ runs over the linear combinations. The constants $\kappa_{a,\lambda}$ and $\kappa_{a,i}$ in \eqref{ortholincomb} are chosen so that $\langle\tilde\CO_{a+\dot+}(x)J^{11}_{+\dot+}(0)\rangle=\langle\tilde\CO_{a+\dot+}(x)\tilde\CO_{b+\dot+}(0)\rangle=0$ at $g=0$. As in the $N=1$ case, we can again check that all $N$ of the operators in \eqref{ortholincomb} must pair up with linearly independent combinations of the $N$ operators $\CO_i=\mu^I_i\left(\sum_{a=1}^N\mu_{aI}\right)$.

Therefore, to justify \eqref{nonnilp}, we can, in analogy with the $N=1$ discussion in the previous subsection, suppose that for some particular $\CO_i$ there is a minimal $\hat n>0$ such that $\CO_i^{\hat n}=0$ in the chiral ring of the collection of the isolated $\CT_i$. Clearly, this operator is in a long multiplet of the following type (the case in which this operator identically vanishes can be ruled out as in the case of a single isolated SCFT described in the previous section)
\eqn{
\hat\CC_{2\hat n-2(0,0)}\oplus\CD_{2\hat n-1(0,0)}\oplus\bar\CD_{2\hat n-1(0,0)}\oplus\hat\CB_{2\hat n}~.
}[hatnlongii]
Now, we study the operator $\CO_{i}^{\hat n-1}\tilde\CO'_{+\dot+}$ in the theory with $g\ne0$, where $\tilde\CO'_{+\dot+}$ is the appropriate linear combination of the $\tilde\CO_{a+\dot+}$ in \eqref{ortholincomb} that pairs up with $\CO_i$. From this discussion, we have
\eqn{
\CO_{i}^{\hat n-1}\tilde\CO'_{+\dot+}\oplus\CO^{\hat n-1}_{i}\tilde\CO'_+\oplus\CO^{\hat n-1}_{i}\tilde\CO'_{\dot+}\oplus\CO_{i}^{\hat n}~.
}[shortlongcontii]
As in the $N=1$ case, we again find a contradiction since $\CO_{i}^{\hat n}$ is already in a long multiplet at $g=0$. In particular, we conclude that 
\eqn{
\CO_{i}^{n}\ne0~,\ \ \  \forall n>0, i=1,\cdots, N~,
}[nonnulliicond]
in the chiral ring of the collection of the $\CT_{i}$. Therefore, as discussed below \eqref{nonnilp}, we see that $|\chi|\ge3$.

\newsec{An example with $|\chi|=3$}

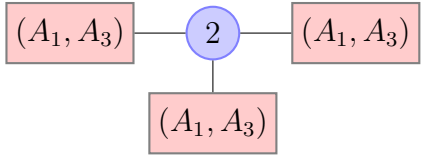
\begin{figure}
\begin{center}
\vskip .6cm
\begin{tikzpicture}[circ/.style={circle,draw=blue!50,fill=blue!20,thick,inner sep=0pt,minimum size=7mm},box/.style={rectangle,draw=black!50,fill=red!20,thick,inner sep=0pt,minimum size=8mm},auto]
\node[box] (1) at (1.1,0) {\;$(A_1, A_3)$\;};
\node[circ] (2) at (3,0) [shape=circle] {$2$} edge [-] node[auto]{} (1);
\node[box] (3) at (4.9,0)  {\;$(A_1, A_3)$\;} edge [-] node[auto]{} (2);
\node[box] (9) at (3,-1.2) {\;$(A_1, A_3)$\;} edge[-] (2);
\end{tikzpicture}
\caption{A quiver diagram describing the theory $\hat{\CT}$ near a weak coupling point on the conformal manifold. The middle circle stands for an $SU(2)$ vector multiplet gauging the diagonal $SU(2)$ flavor symmetry of three $(A_1,A_3)$ theories.}
\label{fig:quiver}
\end{center}
\end{figure}

Given the above general discussion, it is interesting to ask if there are examples that saturate the bound \eqref{bound}. In this section, we will see the answer is yes by explicitly constructing an $\mathcal{N}=2$ SCFT, $\hat{\mathcal{T}}$, that has a one complex-dimensional conformal manifold and whose chiral algebra, $\chi(\hat{\mathcal{T}})$, saturates the bound \eqref{bound}.\foot{We do not consider any hypothetical isolated points on the conformal manifold at finite Zamolodchikov distance where new Schur operators might appear. Such points would have an accidental enhancement of the corresponding chiral algebras. No such points are known to exist in the literature, and they do not affect our discussion at generic points.}

The theory $\hat{\mathcal{T}}$ is given by the low-energy limit of type IIB string theory on the three complex-dimensional hypersurface singularity $x_1^3+x_2^3+x_3^3+\alpha x_1x_2x_3+w^2=0$ \cite{DelZotto:2015rca}, or, alternatively, by three $(A_1,A_3)$ Arygres-Douglas theories whose diagonal $SU(2)$ flavor symmetry is gauged (Fig.~\ref{fig:quiver}). The central charge of the $SU(2)$ flavor symmetry of the $(A_1, A_3)$ theory \cite{Aharony:2007dj} is precisely the right value so that this diagonal gauging is exactly marginal.

The resulting $\mathcal{N}=2$ SCFT, $\hat{\mathcal{T}}$, has conformal anomalies $a_{\hat{\mathcal{T}}} = c_{\hat{\mathcal{T}}} = 2$, and therefore its two-dimensional chiral algebra, $\chi(\hat{\mathcal{T}})$, has 
\begin{align}
c_{2d} = -12 c_{\hat{\mathcal{T}}} = -24~.
\end{align}
In order to identify this chiral algebra, let us compute the Schur index of $\hat{\mathcal{T}}$. Since it is invariant under the exactly marginal gauge coupling, the index can be computed in the weak gauge coupling limit by considering the following integral
\begin{align}
\mathcal{I}_{\hat{\mathcal{T}}}(q) = \oint_{|z|=1}\frac{dz}{2\pi iz}\Delta(z)\,\mathcal{I}_v(q;z)\prod_{i=1}^3\mathcal{I}_{(A_1,A_3)}(q;z)~,
\label{eq:index-hat-T}
\end{align}
where $\Delta(z) \equiv \frac{1}{2}(1-z^2)(1-z^{-2})$ is the Haar measure of $SU(2)$, and 
\eqn{
\mathcal{I}_v(q;z) = P.E.\left[\frac{-2q}{1-q}(z^2 + 1 + z^{-2})\right]~,
}[Ivdef]
is the Schur index of a single $SU(2)$ vector multiplet (recall the definition of the plethystic exponential in \eqref{PEdefn}). The Schur index of the $(A_1,A_3)$ theory was found in \cite{Buican:2015ina} to be\footnote{See also \cite{Cordova:2015nma} for an equivalent expression for the same quantity.}
\begin{align}
\mathcal{I}_{(A_1,A_3)}(q;z) = \frac{1}{\prod_{k=2}^\infty (1-q^k)}\sum_{R}\,[\text{dim}\,R]_q\,\tilde{f}^{(3)}_R(q;z)~,
\end{align}
where $R$ runs over irreducible representations of $su(2)$, and $[k]_q = (q^{\frac{k}{2}} -q^{-\frac{k}{2}})/(q^{\frac{1}{2}}-q^{-\frac{1}{2}})$ for any integer $k$. The factor, $\tilde{f}^{(3)}_R(q;z)$, is given by\footnote{Here, we use a slightly different notation for the flavor fugacity from \cite{Buican:2015ina}. In particular, $x$ in Eq.~(1.6) of \cite{Buican:2015ina} is related to $z$ in this paper by $x=z^2$. In terms of $z$, the character of the $n$-dimensional representation of $su(2)$ is given by $\chi_n(z)  = (z^n-z^{-n})/(z-z^{-1})$. See also footnote 20 of \cite{Buican:2015ina}.}
\begin{align}
\tilde{f}^{(3)}_{R}(q;z) = \frac{q^{3C_2(R)}}{\prod_{k=1}^\infty(1-q^k)} \text{Tr}_{R}(z^{4J_3} q^{-3(J_3)^2})~,
\end{align}
where $J_3$ is the Cartan generator of $su(2)$ normalized so that the fundamental representation has eigenvalues $\pm \frac{1}{2}$, and $C_2(R)$ is the quadratic Casimir invariant.

Now, by an explicit calculation, we see that the index \eqref{eq:index-hat-T} has the following expansion in $q$
\begin{align}
\mathcal{T}_{\hat{\mathcal{T}}}(q) = 1 + q^2 + q^3 + 2q^6 + q^8 + q^{11} + 2q^{12} + q^{15} + 2q^{18} + 2q^{20} + q^{24} + q^{26} + 2q^{27} + \cdots~.
 \label{eq:q-series}
\end{align}
Interestingly, this $q$-series can be written in closed form as
\begin{align}
\mathcal{I}_{\hat{\mathcal{T}}}(q) =  \left(\prod_{k=1}^\infty\frac{1}{1-q^k}\right)\sum_{n=1}^\infty (-1)^{n-1}n\left(q^{\frac{3n^2-n-2}{2}} - q^{\frac{3n^2+n-2}{2}}\right)~.
\label{eq:resum}
\end{align} 
We have checked the equivalence of \eqref{eq:q-series} and \eqref{eq:resum} up to very high perturbative order in $q$. Moreover, the expression \eqref{eq:resum} has no essential singularity at $q\to 1$, which is consistent with $a_{\hat{\CT}}=c_{\hat{\CT}}$.\footnote{The absence of the essential singularity can be seen by rewriting \eqref{eq:resum} using (the derivative of) Jacobi's triple product identity.}

To identify $\chi(\hat\CT)$, our strategy will be to show that the expression in \eqref{eq:resum} coincides with the vacuum character of the  $\mathcal{A}(6)$ algebra of Feigin, Feigin, and Tipunin \cite{Feigin:2007sp, Feigin:2008sg}. The $\mathcal{A}(6)$ algebra is a $\CW$ algebra obtained by extending the Virasoro algebra with central charge $c=-24$ by adding two Virasoro primaries, $\Phi^\pm(z)$, of holomorphic dimension $4$. Since $\Phi^\pm (z)\Phi^\pm (w)$ are anti-symmetric under $z\leftrightarrow w$, $\Phi^\pm$ are fermionic generators. Moreover, $\Phi^\pm$ satisfy the following null state equations 
\begin{align}
\partial^2\Phi^\pm  + \alpha\, T\Phi^\pm = 0~,
\end{align}
where $\alpha$ is a constant, and $T\Phi^\pm$ is the normal ordered product of the stress tensor, $T$, and $\Phi^\pm$.
  The singular part of the operator product expansion of $\Phi^+(z)\Phi^-(w)$ is given by composite operators of the stress tensor and its derivatives. The vacuum character of this algebra is evaluated in \cite{Feigin:2007sp, Feigin:2008sg} as
\begin{align}
Z_{\mathcal{A}(6)}(q,z) \equiv \text{Tr}_{\mathcal{A}(6)}\; q^{L_0} z^{2A}
= \frac{1}{\prod_{n=1}^\infty(1-q^n)}\sum_{n=1}^\infty\sum_{j=-\frac{n-1}{2}}^{\frac{n-1}{2}}z^{2j}\left(q^{\frac{3n^2-n-2}{2}} - q^{\frac{3n^2+n-2}{2}}\right)~,
\end{align}
where $A$ is an $s\ell(2)$ charge under which $\Phi^\pm$ have charge $\pm \frac{1}{2}$, and the stress tensor is neutral.
Note here that, when we set $z = -1$, this vacuum character is equivalent to the expression \eqref{eq:resum}, namely
\begin{align}
\mathcal{I}_{\hat{\mathcal{T}}}(q) = \text{Tr}_{\mathcal{A}(6)} (-1)^{2A}q^{L_0} = Z_{\mathcal{A}(6)}(-1,q)~.
\label{eq:equality}
\end{align} 
Since the $\Phi^\pm$ are fermionic operators with charges $2A=\pm 1$, and $T(z)$ is a bosonic operator with charge $2A=0$, the factor $(-1)^{2A}$ in the trace is identified with $(-1)^F$.\foot{From the four-dimensional perspective, we can think of the $z$ fugacity as being related to a fugacity for rotations in the plane normal to the chiral algebra.} Then, the equality \eqref{eq:equality} is precisely of the form of \eqref{torusI}. 
Moreover, the central charge $c=-24$ of the $\mathcal{A}(6)$ algebra agrees with the expected value of the central charge, $c_{2d}=-12c_{4d}$, of $\chi(\hat{\mathcal{T}})$.
These facts strongly suggest that indeed
\eqn{
\chi(\hat{\mathcal{T}})=\CA(6)~,
}[chiid]
and that therefore, as promised,
\eqn{
|\chi(\hat\CT)|=3~.
}[sat]
\subsec{Explicit check of multiplet recombination in four dimensions}
In this sub-section we further check the identification in \eqref{chiid} by explicitly constructing the Schur operators, up to $\CO(q^4)$, that survive for small but non-zero gauge coupling.\foot{We expect the resulting chiral algebra is the chiral algebra at generic points on the resulting conformal manifold.} As we will see, all chiral algebra generators of $\CO(q^4)$ and lower pair up to form long multiplets, with the exception of the the $SU(2)_R$ current (at $\CO(q^2)$) and the two fermionic generators (at $\CO(q^4)$). Moreover, it is straightforward to check, using the Macdonald index of the $(A_1, A_3)$ theory \cite{Buican:2015tda} and the Macdonald analog of \eqref{eq:index-hat-T}, that the four-dimensional fermionic primaries corresponding to $\Phi^{\pm}$ should be of type $\hat\CC_{{3\over2}({1\over2},0)}\oplus\hat\CC_{{3\over2}(0,{1\over2})}$ (the corresponding Schur operators have charge $\pm{1\over2}$ under $M^{\perp}=j_1-j_2$, which is in turn mapped to charge $\pm{1\over2}$ under $A$ in the chiral algebra).

To check the above statements, we first consider the $\CO(q^2)$ Schur operators at $g=0$
\begin{eqnarray}\label{q2operators}
&&\CO_{ab}=\mu_a^{I}\mu_{bI}\in\hat\CB_2~, \ \ \ \CO_{+a}=\lambda^{1I}_+\mu_{aI}\in\bar\CD_{1(0,0)}~, \ \ \ \CO_{\dot+a}={\rm Tr}\tilde\lambda_{2\dot+}^I\mu_{aI}\in\CD_{1(0,0)}~, \nonumber\\ &&\CO_{+\dot+}=\lambda^{1I}_+\tilde\lambda_{2\dot+I}\in\hat\CC_{0(0,0)}~, \ \ \ J_{a+\dot+}^{11}\in\hat\CC_{0(0,0)}~,
\end{eqnarray}
where $a,b=1,2, 3$ denote the particular $(A_1, A_3)$ factor. Note that the Joseph ideal constraint in each $(A_1, A_3)$ theory implies that $\CO_{aa}=0$. As a result, this constraint leaves three $\CO_{ab}$ operators with $a\ne b$. Using the variations in \eqref{lambdadef}, it straightforward to check that when we turn on a small gauge coupling, all operators in \eqref{q2operators} pair up via
\eqn{
\hat\CC_{0(0,0)}\oplus\CD_{1(0,0)}\oplus\bar\CD_{1(0,0)}\oplus\hat\CB_2~,
}[stressrecomb]
with the exception of the overall $SU(2)_R$ Schur operator that is preserved along the conformal manifold
\eqn{
J_{+\dot+}^{11}=\CO_{+\dot+}-\sum_{a=1}^3J_{a+\dot+}^{11}~.
}

Next, at $\CO(q^3)$, we have the following Schur generators for $g=0$
\begin{eqnarray}\label{q3operators}
&&\CO_{+++}=f_{IJK}\lambda^{1I}_+\lambda^{1J}_+\lambda^{1K}_+\in\bar\CD_{1(1,0)}~, \ \ \ \CO_{\dot+++}=f_{IJK}\lambda^{1I}_+\lambda^{1J}_+\tilde\lambda_{2\dot+}^K\in\hat\CC_{{1\over2}({1\over2},0)}~, \nonumber\\ &&\CO_{\dot+\dot++}=f_{IJK}\tilde\lambda_{2\dot+}^I\tilde\lambda_{2\dot+}^J\lambda^{1K}_+\in\hat\CC_{{1\over2}(0,{1\over2})}~, \ \ \ \CO_{\dot+\dot+\dot+}=f_{IJK}\tilde\lambda_{2\dot+}^I\tilde\lambda_{2\dot+}^J\tilde\lambda_{2\dot+}^K\in\CD_{1(0,1)}~, \nonumber \\&&\CO_{++a}=f_{IJK}\lambda^{1I}_+\lambda^{1J}_+\mu_a^K\in\bar\CD_{{3\over2}({1\over2},0)}~, \ \ \  \CO_{\dot++a}=f_{IJK}\lambda_+^{1I}\tilde\lambda_{2\dot+}^J\mu_a^K\in\hat\CC_{1(0,0)}~, \nonumber\\ && \CO_{\dot+\dot+a}=f_{IJK}\tilde\lambda_{2\dot+}^I\tilde\lambda_{2\dot+}^J\mu_a^K\in\CD_{{3\over2}(0,{1\over2})}~, \ \ \ \CO_{+ab}=f_{IJK}\lambda_+^{1I}\mu_a^J\mu_b^K\in\CD_{2(0,0)}~, \nonumber \\&&\CO_{\dot+ab}=f_{IJK}\tilde\lambda_{2\dot+}^I\mu_a^J\mu_b^K\in\bar\CD_{2(0,0)}~, \ \ \ \CO_{abc}=f_{IJK}\mu_a^I\mu_b^J\mu_c^K\in\hat\CB_3~,\\&&\CO_{\dot++++}=\lambda^{1I}_+D_{+\dot+}\lambda^{1I}_+\in \hat\CC_{0(1,0)}~,\ \ \ \CO_{\dot+\dot+++}=\lambda_+^{1I}D_{+\dot+}\tilde\lambda_{2\dot+I}\in\hat\CC_{0({1\over2},{1\over2})}~, \nonumber\\&& \CO_{\dot+\dot+\dot++}=\tilde\lambda_{2\dot+}^ID_{+\dot+}\tilde\lambda_{2\dot+I}\in\hat\CC_{0(0,1)}~, \ \ \ \CO_{\dot++[ab]}=\mu_{[a}^ID_{+\dot+}\mu_{b]I}\in\hat\CC_{1(0,0)}~, \nonumber\\&&\CO_{\dot+++a}=\lambda^{1I}_+D_{+\dot+}\mu_{aI}\in\hat\CC_{{1\over2}({1\over2},0)}~, \ \ \ \CO_{\dot+\dot++a}=\tilde\lambda_{2\dot+}^ID_{+\dot+}\mu_{aI}\in\hat\CC_{{1\over2}(0,{1\over2})}~,\nonumber
\end{eqnarray}
where $f_{IJK}$ are the $SU(2)$ structure constants. We see that anti-symmetry implies that $\CO_{abc}\sim\CO_{123}$. Similarly, the $\CO_{+ab}$ and $\CO_{\dot+ab}$ operators are anti-symmetric in $a$ and $b$. As a result, there are sixteen fermionic and sixteen bosonic Schur generators at $g=0$.  It is straightforward to check that when we turn on a small but non-zero gauge coupling, all of these operators pair up via
\begin{eqnarray}\label{recombruls}
&&\hat\CC_{1(0,0)}\oplus\CD_{2(0,0)}\oplus\bar\CD_{2(0,0)}\oplus\hat\CB_3~,\nonumber\\ && \hat\CC_{0(1,0)}\oplus\hat\CC_{{1\over2}({1\over2},0)}\oplus\bar\CD_{1(1,0)}\oplus\bar\CD_{{3\over2}({1\over2},0)}~,\nonumber\\ && \hat\CC_{0(0,1)}\oplus\hat\CC_{{1\over2}(0,{1\over2})}\oplus\CD_{1(0,1)}\oplus\CD_{{3\over2}(0,{1\over2})}~,\\&& \hat\CC_{{1\over2}({1\over2},0)}\oplus\hat\CC_{1(0,0)}\oplus\bar\CD_{{3\over2}({1\over2},0)}\oplus\bar\CD_{2(0,0)}~,\nonumber\\&& \hat\CC_{{1\over2}(0,{1\over2})}\oplus\hat\CC_{1(0,0)}\oplus\CD_{{3\over2}(0,{1\over2})}\oplus\CD_{2(0,0)}\nonumber~,\\ \nonumber &&\hat\CC_{0({1\over2},{1\over2})}\oplus\hat\CC_{{1\over2}(0,{1\over2})}\oplus\hat\CC_{{1\over2}({1\over2},0)}\oplus\hat\CC_{1(0,0)}~.
\end{eqnarray}
In particular, there are no new chiral algebra generators at $\CO(q^3)$ for small but non-zero gauge coupling, and the only contribution to the Schur index at this order comes from the descendant, $\partial_{+\dot+}J^{11}_{+\dot+}$.

Finally, let us discuss the $\CO(q^4)$ Schur generators at $g=0$ 
\begin{eqnarray}\label{q4operators}
&&\CO_{\dot+\dot++++}=f_{IJK}\tilde\lambda_{2\dot+}^ID_{+\dot+}\lambda^{1J}_+\lambda^{1K}_+\in\hat\CC_{{1\over2}(1,{1\over2})}~, \CO_{++\dot+\dot+\dot+}=f_{IJK}\lambda^{1I}_+D_{+\dot+}\tilde\lambda_{2\dot+}^J\tilde\lambda_{2\dot+}^K\in\hat\CC_{{1\over2}({1\over2},1)}~, \nonumber\\ &&\CO_{\dot+\dot+\dot++++}=\lambda^{1I}_{+}D_{+\dot+}^2\tilde\lambda_{2\dot+I}\in \ \hat\CC_{0(1,1)}~,\nonumber \ \ \ \CO_{+++\dot+\dot+a}=D_{+\dot+}^2\mu_a^I\lambda^1_{+I}\in\hat\CC_{{1\over2}(1,{1\over2})}~, \nonumber\\&&\CO_{\dot+\dot+\dot+++a}=D_{+\dot+}^2\tilde\lambda_{2\dot+}^I\mu_{aI}\in\hat\CC_{{1\over2}({1\over2},1)}~, \ \ \ \CO_{\dot++++a}=f_{IJK}D_{+\dot+}\lambda_+^{1I}\lambda_{+}^{1J}\mu_a^K\in \hat\CC_{{1}(1,0)}~, \\\nonumber&&\CO_{+\dot+\dot+\dot+a}=f_{IJK}D_{+\dot+}\tilde\lambda^I_{2\dot+}\tilde\lambda^J_{2\dot+}\mu_a^K\in \hat\CC_{{1}(0,1)}~,\ \ \ \CO_{\dot+\dot+++a}=f_{IJK}D_{+\dot+}\lambda_+^{1I}\tilde\lambda^J_{2\dot+}\mu_a^K~, \\&&\CO'_{\dot+\dot+++a}=f_{IJK}\lambda_+^{1I}D_{+\dot+}\tilde\lambda_{2\dot+}^J\mu_a^K\in\hat\CC_{{1({1\over2},{1\over2})}}~,\nonumber\ \ \ \CO_{++\dot+\dot+ab}=D_{+\dot+}^2\mu_a^I\mu_{bI}\in\hat\CC_{1({1\over2},{1\over2})}~, \\&& \CO_{++\dot+ab}=f_{IJK}\lambda^{1I}_+\mu_a^JD_{+\dot+}\mu_b^K\in\hat\CC_{{3\over2}({1\over2},0)}~, \ \ \ \CO_{\dot+\dot++ab}=f_{IJK}\tilde\lambda_{2\dot+}^I\mu_a^JD_{+\dot+}\mu_b^K\in\hat\CC_{{3\over2}(0,{1\over2})}~,\nonumber\\&&\CO_{+\dot+abc}=f_{IJK}\mu_a^I\mu_b^JD_{+\dot+}\mu_c^K\in\hat\CC_{2(0,0)}~.\nonumber
\end{eqnarray}

Naively, we have twenty-six fermionic generators and twenty-seven bosonic generators. However, six bosonic and six fermionic constraints come from applying (4.15) of \cite{Buican:2015tda} in each $(A_1, A_3)$ sector. In particular, we have
\begin{eqnarray}\label{6cons}
&&J^{11}_{a+\dot+}\CO_{ab}=J^{11}_{a+\dot+}{\rm Tr}[\mu_a\mu_b]={1\over\sqrt{2}}f_{ABC}\mu_b^A\mu_a^BD_{+\dot+}\mu_a^C\sim\CO_{+\dot+baa}\nonumber~,\\&&J^{11}_{a+\dot+}\CO_{+a}=J^{11}_{a+\dot+}{\rm Tr}[\mu_a\lambda^1_+]={1\over\sqrt{2}}f_{ABC}(\lambda_+^1)^A\mu_a^BD_{+\dot+}\mu_a^C\sim\CO_{++\dot+aa}~,\\&&J^{11}_{a+\dot+}\CO_{\dot+a}=J^{11}_{a+\dot+}{\rm Tr}[\mu_a\tilde\lambda_{2\dot+}]={1\over\sqrt{2}}f_{ABC}(\tilde\lambda_{2\dot+})^A\mu_a^BD_{+\dot+}\mu_a^C\sim\CO_{\dot+\dot++aa}~.\nonumber
\end{eqnarray}
Note that the case $a=b$ is not a new constraint (indeed this follows from the Joseph ideal constraints, $\CO_{aa}=0$). The final three bosonic relations can be seen in the form of the Macdonald index in \rcite{Buican:2015tda}. In particular, according to table 1 of \rcite{Buican:2015tda}, there is only one singlet of $SU(2)$ at $O(q^2t^2)$. The Joseph ideal constraint eliminates $\partial^2_{+\dot+}(\mu_a^I\mu_{aI})$, which leaves $D_{+\dot+}^2\mu_a^I\mu_{aI}=\CO_{\dot+\dot+++aa}$, and $(J^{11}_{a+\dot+})^2$. Therefore, we must have three relations (one in each $(A_1, A_3)$ sector) of the form
\eqn{
\kappa(J^{11}_{a+\dot+})^2=\CO_{\dot+\dot+++aa}~,
}[newrelat]
where $\kappa$ is a constant. As a result, we are left with twenty fermionic generators and eighteen bosonic generators. When we turn on the gauge coupling, all the bosonic generators pair up with eighteen fermions via
\begin{eqnarray}\label{rrules}
&&\hat\CC_{0(1,1)}\oplus\hat\CC_{{1\over2}(1,{1\over2})}\oplus\hat\CC_{{1\over2}({1\over2},1)}\oplus\hat\CC_{1({1\over2},{1\over2})}~,\nonumber\\&&\hat\CC_{{1\over2}(1,{1\over2})}\oplus\hat\CC_{{1}(1,{0})}\oplus\hat\CC_{{1}({1\over2},{1\over2})}\oplus\hat\CC_{{3\over2}({1\over2},{0})}~,\nonumber\\&&\hat\CC_{{1\over2}({1\over2},1)}\oplus\hat\CC_{{1}(0,{1})}\oplus\hat\CC_{{1}({1\over2},{1\over2})}\oplus\hat\CC_{{3\over2}(0,{1\over2})}~,\\&&\hat\CC_{{1}({1\over2},{1\over2})}\oplus\hat\CC_{{3\over2}({1\over2},{0})}\oplus\hat\CC_{{3\over2}({0},{1\over2})}\oplus\hat\CC_{{2}(0,{0})}~.\nonumber
\end{eqnarray}
It is easy to check that only one linear combination of the $\CO_{++\dot+ab}$ and one linear combination of the $\CO_{\dot+\dot++ab}$ remain un-lifted. These operators are precisely the generators of type $\hat\CC_{{3\over2}({1\over2},0)}\oplus\hat\CC_{{3\over2}(0,{1\over2})}$ we are looking for (they cancel the $\CO(q^4)$ contributions from the stress tensor multiplet: $\partial^2_{+\dot+}J^{11}_{+\dot+}$ and $(J^{11}_{+\dot+})^2$).

\newsec{Discussion and Conclusion}
In this paper, we  found  new connections between the conformal manifold and the moduli space of vacua. We also saw that $\CN=2$ theories with $a\ge c$ must have additional fermionic symmetries in two dimensions (as long as the theories we study satisfy the Cardy-like scaling in \eqref{Schurcardy} and \cite{DiPietro:2014bca}). Finally, we demonstrated that the chiral ring---specifically the subring that controls exactly marginal deformations---and the chiral algebra are interrelated (in the sense that the existence of an exactly marginal gauge coupling imposes \eqref{bound} on the resulting chiral algebras). We suspect that there is a deeper connection and therefore suggest the following open problems:
\begin{itemize}
\item[$\bullet$] All examples of chiral algebras arising from conformal manifolds seem to admit non-trivial actions of certain Lie algebras. In the case of the Lagrangian conformal theories this is the manifest flavor symmetry in the four-dimensional description. On the other hand, in our theory, $\hat\CT$, we had no flavor symmetry in four dimensions. The two dimensional $sl(2)$ symmetry instead arose from the non-trivial action of the rotational symmetry transverse to the chiral algebra plane. It would clearly be desirable to have an explanation (or refutation) of this apparently general phenomenon.
\item[$\bullet$] It follows from our work that if there is an $\CN=2$ theory (satisfying the usual Cardy-like scaling of \eqref{Schurcardy} and \cite{DiPietro:2014bca}) with $|\chi|<3$ and an exactly marginal deformation, then that marginal deformation would be exotic: it would not have an interpretation as a gauge coupling.
\item[$\bullet$] It would be interesting to understand if the equation $\partial^2\Phi^{\pm}+\alpha T\Phi^{\pm}=0$ when acting on fermionic $sl(2)$ doublets, $\Phi^{\pm}$, in chiral algebras arising from four dimensions is enough to guarantee the existence of an exactly marginal deformation in the associated four-dimensional theory. This statement holds in the $\CA(6)$ theory we studied and also in the case of the chiral algebra corresponding to the free vector multiplet.
\item[$\bullet$] Identify the natural mathematical structures associated with chiral algebras coming from conformal manifolds in four dimensions.
\end{itemize}

\ack{ \bigskip
We would like to thank D.~Kutasov, C.~Papageorgakis, S.~Ramgoolam, L.~Rastelli, and R.~Russo for discussions and correspondence on this and related topics. M.~B.'s work is partially supported by the Royal Society under the grant \lq\lq New Constraints and Phenomena in Quantum Field Theory" and by the U.S. Department of Energy under grant DE-SC0009924. T.~N. is partially supported by the Yukawa Memorial Foundation.
}

\newpage
\bibliography{chetdocbib}

\begin{thebibliography}{10}
\ifx\href\asklfhas\newcommand{\href}[2]{#2}\fi
\ifx\arxivref\asklfhas\newcommand{\arxivref}[2]{\href{http://arxiv.org/abs/#1}{#2}}\fi
\ifx\doiref\asklfhas\newcommand{\doiref}[2]{\href{http://dx.doi.org/#1}{#2}}\fi
\parskip 0pt
\normalsize

\bibitem{Seiberg:1994rs}
N.~Seiberg \& E.~Witten,
\textit{``{Electric - magnetic duality, monopole condensation, and confinement
  in N=2 supersymmetric Yang-Mills theory}''},
\doiref{10.1016/0550-3213(94)90124-4}{Nucl.~Phys. \textbf{B426}, 19 (1994)},
\normalsize{\texttt{\arxivref{hep-th/9407087}{hep-th/9407087}}},
[Erratum: Nucl. Phys.B430,485(1994)].

\bibitem{Seiberg:1994aj}
N.~Seiberg \& E.~Witten,
\textit{``{Monopoles, duality and chiral symmetry breaking in N=2
  supersymmetric QCD}''},
\doiref{10.1016/0550-3213(94)90214-3}{Nucl.~Phys. \textbf{B431}, 484 (1994)},
\normalsize{\texttt{\arxivref{hep-th/9408099}{hep-th/9408099}}}.

\bibitem{Beem:2013sza}
C.~Beem, M.~Lemos, P.~Liendo, W.~Peelaers, L.~Rastelli et~al.,
\textit{``{Infinite Chiral Symmetry in Four Dimensions}''},
\doiref{10.1007/s00220-014-2272-x}{Commun.Math.Phys. \textbf{336}, 1359
  (2015)},
\normalsize{\texttt{\arxivref{1312.5344}{arXiv:1312.5344}}}.

\bibitem{Buican:2015ina}
M.~Buican \& T.~Nishinaka,
\textit{``{On the superconformal index of ArgyresŽÐDouglas theories}''},
\doiref{10.1088/1751-8113/49/1/015401}{J.~Phys. \textbf{A49}, 015401 (2016)},
\normalsize{\texttt{\arxivref{1505.05884}{arXiv:1505.05884}}}.

\bibitem{Buican:2015hsa}
M.~Buican \& T.~Nishinaka,
\textit{``{Argyres-Douglas Theories, $S^1$ Reductions, and Topological
  Symmetries}''},
\doiref{10.1088/1751-8113/49/4/045401}{J.~Phys. \textbf{A49}, 045401 (2016)},
\normalsize{\texttt{\arxivref{1505.06205}{arXiv:1505.06205}}}.

\bibitem{Buican:2015tda}
M.~Buican \& T.~Nishinaka,
\textit{``{Argyres-Douglas Theories, the Macdonald Index, and an RG
  Inequality}''},
\normalsize{\texttt{\arxivref{1509.05402}{arXiv:1509.05402}}}.

\bibitem{Cordova:2015nma}
C.~Cordova \& S.H. Shao,
\textit{``{Schur Indices, BPS Particles, and Argyres-Douglas Theories}''},
\doiref{10.1007/JHEP01(2016)040}{JHEP \textbf{1601}, 040 (2016)},
\normalsize{\texttt{\arxivref{1506.00265}{arXiv:1506.00265}}}.

\bibitem{Song:2015wta}
J.~Song,
\textit{``{Superconformal indices of generalized Argyres-Douglas theories from
  2d TQFT}''},
\normalsize{\texttt{\arxivref{1509.06730}{arXiv:1509.06730}}}.

\bibitem{Cecotti:2015lab}
S.~Cecotti, J.~Song, C.~Vafa \& W.~Yan,
\textit{``{Superconformal Index, BPS Monodromy and Chiral Algebras}''},
\normalsize{\texttt{\arxivref{1511.01516}{arXiv:1511.01516}}}.

\bibitem{Buican:2016hnq}
M.~Buican \& T.~Nishinaka,
\textit{``{A Small Deformation of a Simple Theory}''},
\normalsize{\texttt{\arxivref{1602.05545}{arXiv:1602.05545}}}.

\bibitem{Xie:2016hny}
D.~Xie \& K.~Yonekura,
\textit{``{A search for minimal 4d N=1 SCFT}''},
\normalsize{\texttt{\arxivref{1602.04817}{arXiv:1602.04817}}}.

\bibitem{Feigin:2007sp}
B.~Feigin, E.~Feigin \& I.~Tipunin,
\textit{``{Fermionic formulas for (1,p) logarithmic model characters in
  Phi{2,1} quasiparticle realisation}''},
\normalsize{\texttt{\arxivref{0704.2464}{arXiv:0704.2464}}}.

\bibitem{Feigin:2008sg}
B.L. Feigin \& I.{\relax Yu}. Tipunin,
\textit{``{Characters of coinvariants in (1,p) logarithmic models}''},
\normalsize{\texttt{\arxivref{0805.4096}{arXiv:0805.4096}}}.

\bibitem{Cecotti:2010fi}
S.~Cecotti, A.~Neitzke \& C.~Vafa,
\textit{``{R-Twisting and 4d/2d Correspondences}''},
\normalsize{\texttt{\arxivref{1006.3435}{arXiv:1006.3435}}}.

\bibitem{Rastelli}
L.~Rastelli,
\textit{``{ÒInfinite Chiral Symmetry in Four and Six Dimensions}''},
Seminar~at~Harvard~University \textbf{}, November 2014.

\bibitem{DiPietro:2014bca}
L.~Di~Pietro \& Z.~Komargodski,
\textit{``{Cardy formulae for SUSY theories in $d =$ 4 and $d =$ 6}''},
\doiref{10.1007/JHEP12(2014)031}{JHEP \textbf{1412}, 031 (2014)},
\normalsize{\texttt{\arxivref{1407.6061}{arXiv:1407.6061}}}.

\bibitem{Ardehali:2015bla}
A.A. Ardehali,
\textit{``{High-temperature asymptotics of supersymmetric partition
  functions}''},
\normalsize{\texttt{\arxivref{1512.03376}{arXiv:1512.03376}}}.

\bibitem{Nishinaka:2016hbw}
T.~Nishinaka \& Y.~Tachikawa,
\textit{``{On 4d rank-one N=3 superconformal field theories}''},
\normalsize{\texttt{\arxivref{1602.01503}{arXiv:1602.01503}}}.

\bibitem{Gadde:2011uv}
A.~Gadde, L.~Rastelli, S.S. Razamat \& W.~Yan,
\textit{``{Gauge Theories and Macdonald Polynomials}''},
\doiref{10.1007/s00220-012-1607-8}{Commun.~Math.~Phys. \textbf{319}, 147
  (2013)},
\normalsize{\texttt{\arxivref{1110.3740}{arXiv:1110.3740}}}.

\bibitem{Dolan:2002zh}
F.A. Dolan \& H.~Osborn,
\textit{``{On short and semi-short representations for four-dimensional
  superconformal symmetry}''},
\doiref{10.1016/S0003-4916(03)00074-5}{Annals~Phys. \textbf{307}, 41 (2003)},
\normalsize{\texttt{\arxivref{hep-th/0209056}{hep-th/0209056}}}.

\bibitem{Dobrev:1985qv}
V.K. Dobrev \& V.B. Petkova,
\textit{``{All Positive Energy Unitary Irreducible Representations of Extended
  Conformal Supersymmetry}''},
\doiref{10.1016/0370-2693(85)91073-1}{Phys.~Lett. \textbf{B162}, 127 (1985)}.

\bibitem{Xie:2012hs}
D.~Xie,
\textit{``{General Argyres-Douglas Theory}''},
\doiref{10.1007/JHEP01(2013)100}{JHEP \textbf{1301}, 100 (2013)},
\normalsize{\texttt{\arxivref{1204.2270}{arXiv:1204.2270}}}.

\bibitem{Xie:2013jc}
D.~Xie \& P.~Zhao,
\textit{``{Central charges and RG flow of strongly-coupled N=2 theory}''},
\doiref{10.1007/JHEP03(2013)006}{JHEP \textbf{1303}, 006 (2013)},
\normalsize{\texttt{\arxivref{1301.0210}{arXiv:1301.0210}}}.

\bibitem{Buican:2014hfa}
M.~Buican, S.~Giacomelli, T.~Nishinaka \& C.~Papageorgakis,
\textit{``{Argyres-Douglas Theories and S-Duality}''},
\doiref{10.1007/JHEP02(2015)185}{JHEP \textbf{1502}, 185 (2015)},
\normalsize{\texttt{\arxivref{1411.6026}{arXiv:1411.6026}}}.

\bibitem{DelZotto:2015rca}
M.~Del~Zotto, C.~Vafa \& D.~Xie,
\textit{``{Geometric engineering, mirror symmetry and $
  6{\mathrm{d}}_{\left(1,0\right)}\to
  4{\mathrm{d}}_{\left(\mathcal{N}=2\right)} $}''},
\doiref{10.1007/JHEP11(2015)123}{JHEP \textbf{1511}, 123 (2015)},
\normalsize{\texttt{\arxivref{1504.08348}{arXiv:1504.08348}}}.

\bibitem{Hanany:2010qu}
A.~Hanany \& N.~Mekareeya,
\textit{``{Tri-vertices and SU(2)'s}''},
\doiref{10.1007/JHEP02(2011)069}{JHEP \textbf{1102}, 069 (2011)},
\normalsize{\texttt{\arxivref{1012.2119}{arXiv:1012.2119}}}.

\bibitem{DiFrancesco:1997nk}
P.~Di~Francesco, P.~Mathieu \& D.~Senechal,
\textit{``{Conformal Field Theory}''},
Springer-Verlag (1997),
New York.

\bibitem{Aharony:2007dj}
O.~Aharony \& Y.~Tachikawa,
\textit{``{A Holographic computation of the central charges of d=4, N=2
  SCFTs}''},
\doiref{10.1088/1126-6708/2008/01/037}{JHEP \textbf{0801}, 037 (2008)},
\normalsize{\texttt{\arxivref{0711.4532}{arXiv:0711.4532}}}.

\end{thebibliography}

\end{document}